\documentclass[twocolumn,pre,aps,floatfix]{revtex4-1}

\usepackage{amsmath}
\usepackage{graphicx}
\usepackage{epstopdf}
\usepackage[usenames]{color}
\usepackage{xr}
\usepackage[breaklinks=true,
            pdfborder={0 0 1},
            colorlinks=true,
            linkcolor=black,
            citecolor=blue,
            urlcolor=blue]{hyperref}
\usepackage[normalem]{ulem}
\usepackage{rotating}
\usepackage{verbatim}
\usepackage{bibentry}

\def\Dr{D_\text{r}}
\def\kt{k_\text{B}T}
\def\Pe{\text{Pe}}
\def\vp{v_\text{p}}
\def\fc{f_\text{c}}
\def\kin{k_\text{in}}
\def\kout{k_\text{out}}
\def\rhog{\rho_\text{g}}
\def\rhoc{\rho_\text{c}}
\def\Fmax{F_\text{max}}
\def\fmax{f_\text{max}}
\def\nuhat{\hat{\boldsymbol{\nu}}}
\def\nhat{\hat{\boldsymbol{n}}}

\begin{document}

\title{Reentrant Phase Behavior in Active Colloids with Attraction}

\author{Gabriel S. Redner}
\author{Aparna Baskaran}
\author{Michael F. Hagan}
\email{hagan@brandeis.edu}

\affiliation{Martin Fisher School of Physics, Brandeis University,
  Waltham, MA, USA.}

\begin{abstract}
Motivated by recent experiments, we study a system of self-propelled
colloids that experience short-range attractive interactions and are
confined to a surface.  Using simulations we find that the phase
behavior for such a system is reentrant as a function of activity:
phase-separated states exist in both the low- and high-activity
regimes, with a homogeneous active fluid in between.  To understand
the physical origins of reentrance, we develop a kinetic model for the
system's steady-state dynamics whose solution captures the main
features of the phase behavior.  We also describe the varied kinetics
of phase separation, which range from the familiar nucleation and
growth of clusters to the complex coarsening of active particle gels.
\end{abstract}

\maketitle

\section{Introduction}

The collective behaviors of swarming organisms such as birds, fish,
insects, and bacteria have long been subjects of wonder and
fascination, as well as scientific study \cite{Vicsek2012}.  From a
physicist's perspective, such systems can be understood as fluids
driven far from equilibrium by the injection of kinetic energy at the
scale of individual particles, leading to a zoo of unusual phenomena
such as dynamical self-regulation \cite{Gopinath2012}, clustering
\cite{Peruani2006, Tailleur2008, Cates2013, Redner2013, Fily2012,
  Buttinoni2013a}, segregation \cite{McCandlish2012}, anomalous
density fluctuations \cite{Search2003}, and strange rheological and
phase behavior \cite{Giomi2010, Saintillan2010, Cates2008, Shen2004,
  Bialke2012}.  Recently, nonliving systems that also exhibit
collective behaviors have been constructed from chemically propelled
particles undergoing self-diffusophoresis \cite{Palacci2013,
  Palacci2010, Paxton2004, Hong2007}, squirming droplets
\cite{Thutupalli2011a}, Janus particles undergoing thermophoresis
\cite{Jiang2010, Volpe2011}, and vibrated monolayers of granular
particles \cite{Narayan2007, Kudrolli2008a, Deseigne2010}, suggesting
the possibility of creating a new class of active materials with
properties not achievable with traditional materials.  However,
designing such systems is presently hindered by an incomplete
understanding of how emergent patterns and dynamics depend on the
interplay between activity and microscopic interparticle interactions.

\begin{figure}[h!]
  \includegraphics[width=\columnwidth]{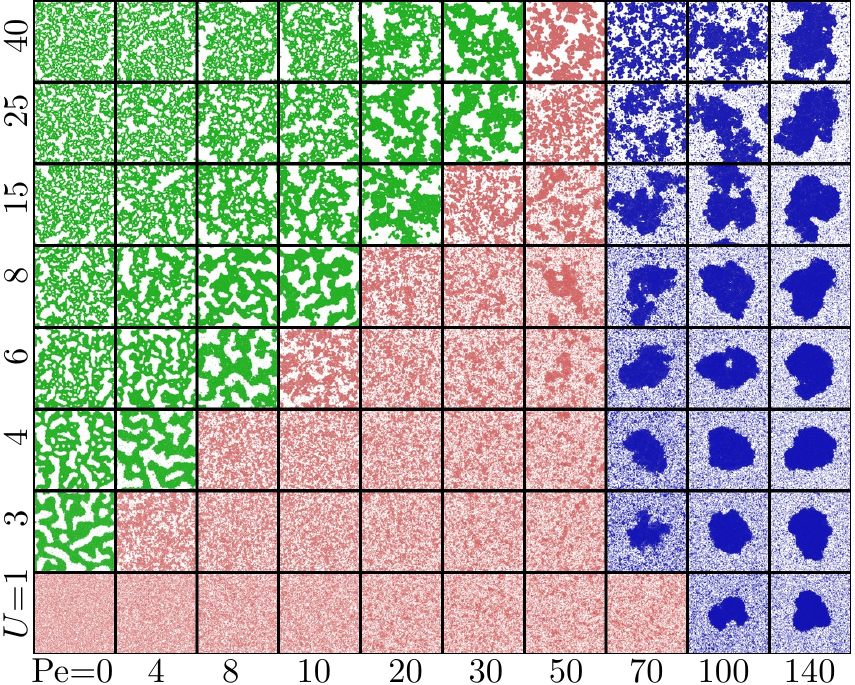}\\
  \includegraphics[width=\columnwidth]{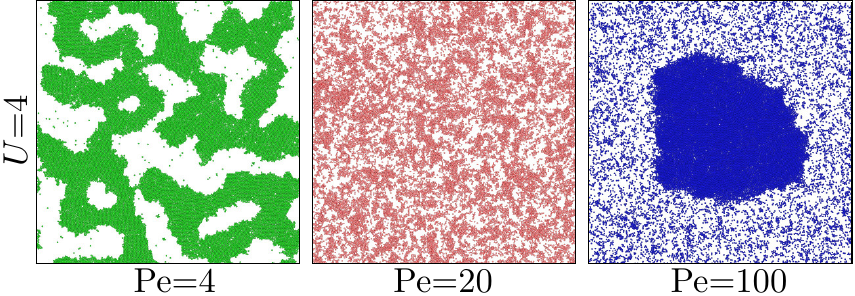}
\caption{(Color online) {\bf Top:} Phase diagram illustrated by
  simulation snapshots at time $t=1000 \tau$ with area fraction $\phi
  = 0.4$, as a function of interparticle attraction strength $U$ and
  propulsion strength $\Pe$.  The colors are a guide to the eye
  distinguishing near-equilibrium gel states (upper-left) from
  single-phase active fluids (center) and self-trapping-induced
  phase-separated states (right).  {\bf Bottom:} Detail of reentrant
  phase behavior at $U = 4$ as $\Pe$ is increased.  At $\Pe=4$ (left),
  the system is nearly thermal and forms a kinetically arrested
  attractive gel.  Increasing $\Pe$ to $20$ (center) suppresses phase
  separation and produces a homogeneous fluid characterized by
  transient density fluctuations.  Increasing $\Pe$ further to $100$
  (right) results in athermal phase separation induced by
  self-trapping.}
  \label{fig:overview}
\end{figure}

In this work we investigate an apparent conflict between two
recently-reported effects of activity on the phase behavior of active
fluids.  We recently studied \cite{Redner2013} a system of
self-propelled hard spheres (with no attractive interactions) in which
activity \emph{induces} a continuous phase transition to a state in
which a high density solid coexists with a low density fluid, complete
with a binodal coexistence curve and critical point.  This athermal
phase separation is driven by self-trapping \cite{Tailleur2008,
  Cates2013}.  Separately, a study of swimming bacteria with
depletion-induced attractive interactions \cite{Schwarz-Linek2012}
demonstrated that activity \emph{suppresses} phase separation, an
effect which those authors postulate is generic.

We resolve this apparent paradox by demonstrating that the phase
diagram for a system of particles endowed with both attractive
interactions and nonequilibrium self-propulsion is {\bf reentrant} as
a function of activity.  Depending on parameter values, activity can
either compete with interparticle attractions to suppress phase
separation or act cooperatively to enhance it.  At low activity the
system is phase-separated due to attraction, while moderate activity
levels suppress this clustering to produce a homogeneous fluid.
Increasing activity even further induces self-trapping which returns
the system to a phase-separated state.  We construct a simple kinetic
model whose analytic solution captures the form of this unusual phase
diagram and explains the mechanism by which activity can both suppress
and promote phase separation in different regimes.  We also describe
the kinetics of phase separation, which differ significantly between
the near-equilibrium and high-activity phase-separated states.  The
behaviors we observe are robust to variations in parameter values, and
thus could likely be observed in experimental systems of
self-propelled, attractive colloids such as those studied in
Refs. \cite{Theurkauff2012a, Schwarz-Linek2012, Palacci2013}.

\section{Model}

Our model is motivated by recently developed experimental systems of
self-propelled colloids sedimented at an interface
\cite{Theurkauff2012a, Palacci2013}, and consists of smooth spheres
immersed in a solvent and confined to a two-dimensional plane
\cite{Redner2013}.  Each particle is active, propelling itself forward
at a constant speed.  Since the particles are smooth spheres and we
neglect all hydrodynamic coupling \footnote{Neglecting hydrodynamic
  coupling can be justified either by restricting the domain of
  applicability to motile cells on a surface \cite{Fily2012}, by
  noting that hydrodynamic coupling is screened and decays rapidly in
  space for particles close to a hard wall \cite{Diamant2005,
    Dufresne2000}, or by observing as in two recent studies of active
  colloids \cite{Palacci2013, Buttinoni2013a} that the collective
  phenomena observed in experiments can be well reproduced in
  simulations without including hydrodynamic interactions.},
\nocite{Diamant2005} \nocite{Dufresne2000} they do not interchange
angular momentum and thus there are no systematic torques which might
lead to alignment. However, the particles' self-propulsion directions
undergo rotational diffusion; based on experimental observations
\cite{Theurkauff2012a}, we confine the propulsion directions to be
always parallel to the surface. For simplicity, interparticle
interactions are modeled by the standard Lennard-Jones potential
$V_{\mathrm{LJ}} = 4 \epsilon \left[ \left( \frac{\sigma}{r}
  \right)^{12} - \left( \frac{\sigma}{r} \right)^6 \right]$ which
provides hard-core repulsion as well as short-range attraction, with
$\sigma$ the nominal particle diameter, and $\epsilon$ the depth of
the attractive well.

The state of the system is represented by the positions and
self-propulsion directions $\{\boldsymbol{r}_i, \theta_i\}_{i=1}^N$ of
the particles, and their evolution is governed by the coupled
overdamped Langevin equations:
\begin{align}
  \dot{\boldsymbol{r}}_i &= \frac{1}{\gamma}
  \boldsymbol{F}_{\mathrm{LJ}}(\{\boldsymbol{r}_i\}) + \vp
  \nuhat_i + \sqrt{2 D} \, \boldsymbol{\eta}^T_i \\ \dot{\theta}_i
  &= \sqrt{2 \Dr} \, \eta^R_i
\end{align}
Here $\boldsymbol{F}_{\mathrm{LJ}} = - \nabla V_\mathrm{LJ}$, $\vp$ is
the magnitude of the self-propulsion velocity, and $\nuhat_i = (\cos
\theta_i, \sin \theta_i)$.  The Stokes drag coefficient $\gamma$ is
related to the diffusion constant by the Einstein relation $D =
\frac{\kt}{\gamma}$.  $\Dr$ is the rotational diffusion constant,
which for a sphere in the low-Reynolds-number regime is $\Dr =
\frac{3D}{\sigma^2}$. The $\eta$ are Gaussian white noise variables
with $\left\langle \eta_i(t) \right\rangle = 0$ and $\left\langle
\eta_i(t) \eta_j(t') \right\rangle = \delta_{ij} \delta(t-t')$.

We non-dimensionalized the equations of motion using $\sigma$ and
$\kt$ as basic units of length and energy, and $\tau =
\frac{\sigma^2}{D}$ as the unit of time.  Our Brownian dynamics
simulations employed the stochastic Runge-Kutta method
\cite{Branka1999} with an adaptive time step, with maximum value $2
\times 10^{-5} \tau$. The potential $V_\mathrm{LJ}$ was cut off and
shifted at $r=2.5 \sigma$.

\section{Phase Behavior}

We parametrize the system by three dimensionless variables: the area
fraction $\phi$, the P\'eclet number $\Pe = \vp \frac{\tau}{\sigma}$,
and the strength of attraction $U = \frac{\epsilon}{\kt}$.  In order
to limit our investigation to regions with nontrivial phase behavior,
we fix the area fraction at $\phi = 0.4$.  At this density, a passive
system ($\Pe=0$) is supercritical for $U \lesssim 2.2$, and
phase-separated for stronger interactions \cite{Smit1991}.  For purely
repulsive self-propelled particles, the system undergoes athermal
phase separation as a result of self-trapping for $\Pe \gtrsim 85$,
and remains a homogeneous fluid for smaller $\Pe$ \cite{Redner2013}.

To understand the phase behavior away from these limits, we performed
simulations in a periodic box with side length $L=200$ (with resulting
particle count $N = 20371$) for a range of attraction strengths $U \in
[1,50]$ and propulsion strengths $\Pe \in [0,160]$.  Except where
noted, each simulation was run until time $1000 \tau$.  Systems were
initialized with random particle positions and orientations except
that (1) particles were not allowed to overlap and (2) each system
initially contained a close-packed hexagonal cluster comprised of
$1000$ particles to overcome any nucleation barriers.  To quantify
clustering, we consider two particles bonded if their centers are
closer than a threshold, and identify clusters as bonded sets of more
than 200 particles.  The cluster fraction $\fc$ is then calculated as
the total number of particles in clusters divided by $N$.

The behavior of the system is illustrated in Fig. \ref{fig:overview}
by representative snapshots (see also
\ref*{Supplement-fig:ljas-0.40-004-004},
\ref*{Supplement-fig:ljas-0.40-020-004}, and
\ref*{Supplement-fig:ljas-0.40-100-004} in the SI
\cite{SupplementalInformation}), and in Fig. \ref{fig:phase} with a
contour plot of $\fc$.  The most striking result is that the phase
diagram is reentrant as a function of $\Pe$.  As shown in
Fig. \ref{fig:overview}, low-$\Pe$ systems form kinetically arrested
gels \cite{Trappe2004} which gradually coarsen toward bulk phase
separation.  Increasing $\Pe$ to a moderate level destabilizes these
aggregates and produces a homogeneous fluid, while increasing activity
beyond a second threshold accesses a high-$\Pe$ regime in which
self-trapping \cite{Redner2013, Tailleur2008, Cates2013} restores the
system to a phase-separated state.

As evident in Fig. \ref{fig:overview}, the width of the intermediate
single-phase region shrinks as the attraction strength $U$ increases,
eliminating reentrance for $U \gtrsim 40$. This trend can be
schematically understood as follows.  In the low-activity gel states,
particles are reversibly bonded by energetic attraction.  Particles
thus arrested have random orientations, and so the mean effect of
self-propulsion is to break bonds and pull aggregates apart.  This
opposes the influence of attraction, and so the width of the low-$\Pe$
gel region increases with $U$.  By contrast, at high $\Pe$ we find
that self-trapping is the primary driver of aggregation.  As shown in
the next section, energetic attractions act cooperatively with
self-trapping in this regime to enable phase separation at lower $\Pe$
than would be possible with activity alone.

\section{Kinetic Model}

\begin{figure}[tbp]
  \includegraphics[width=.48\columnwidth]{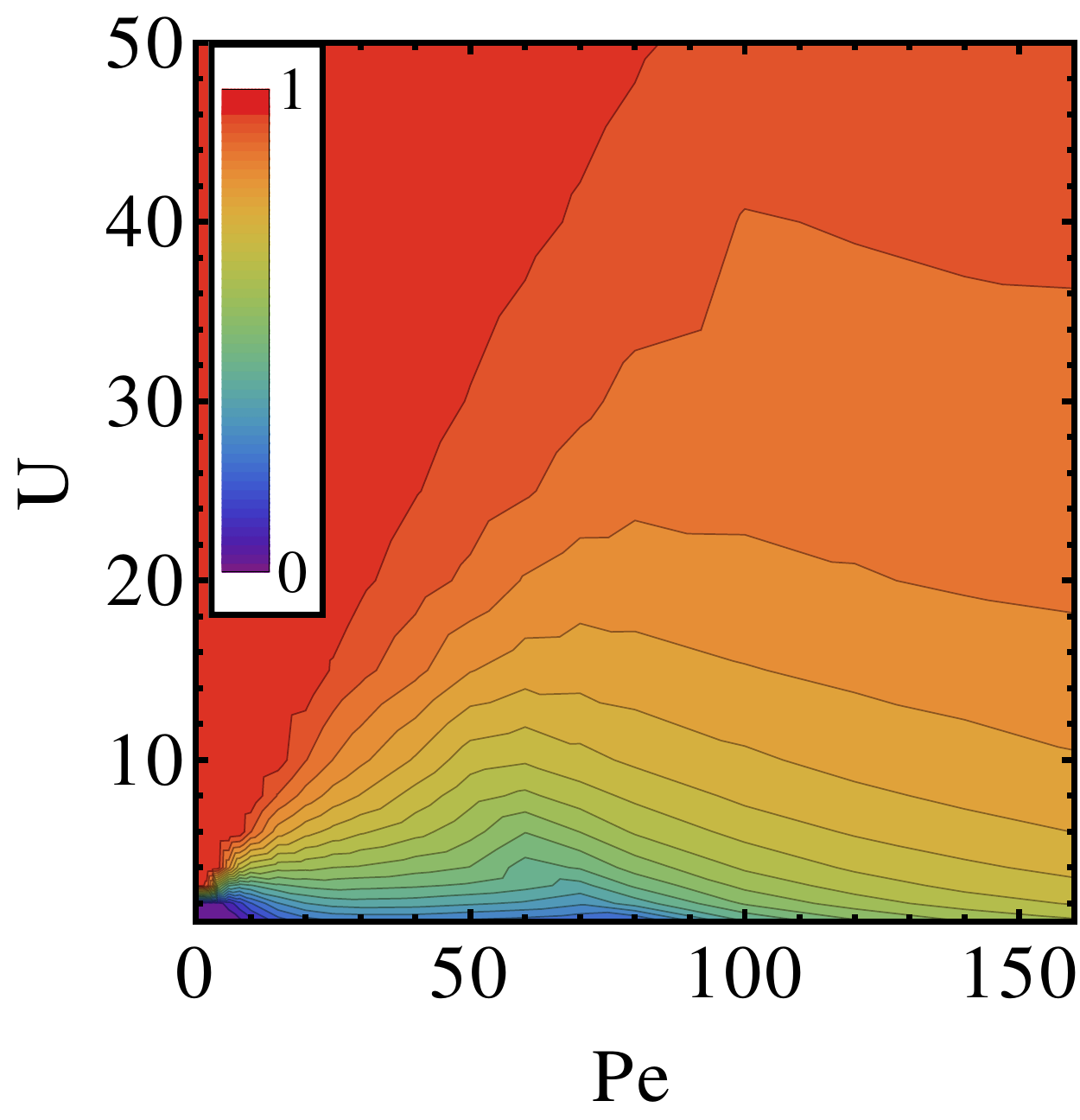}
  \includegraphics[width=.48\columnwidth]{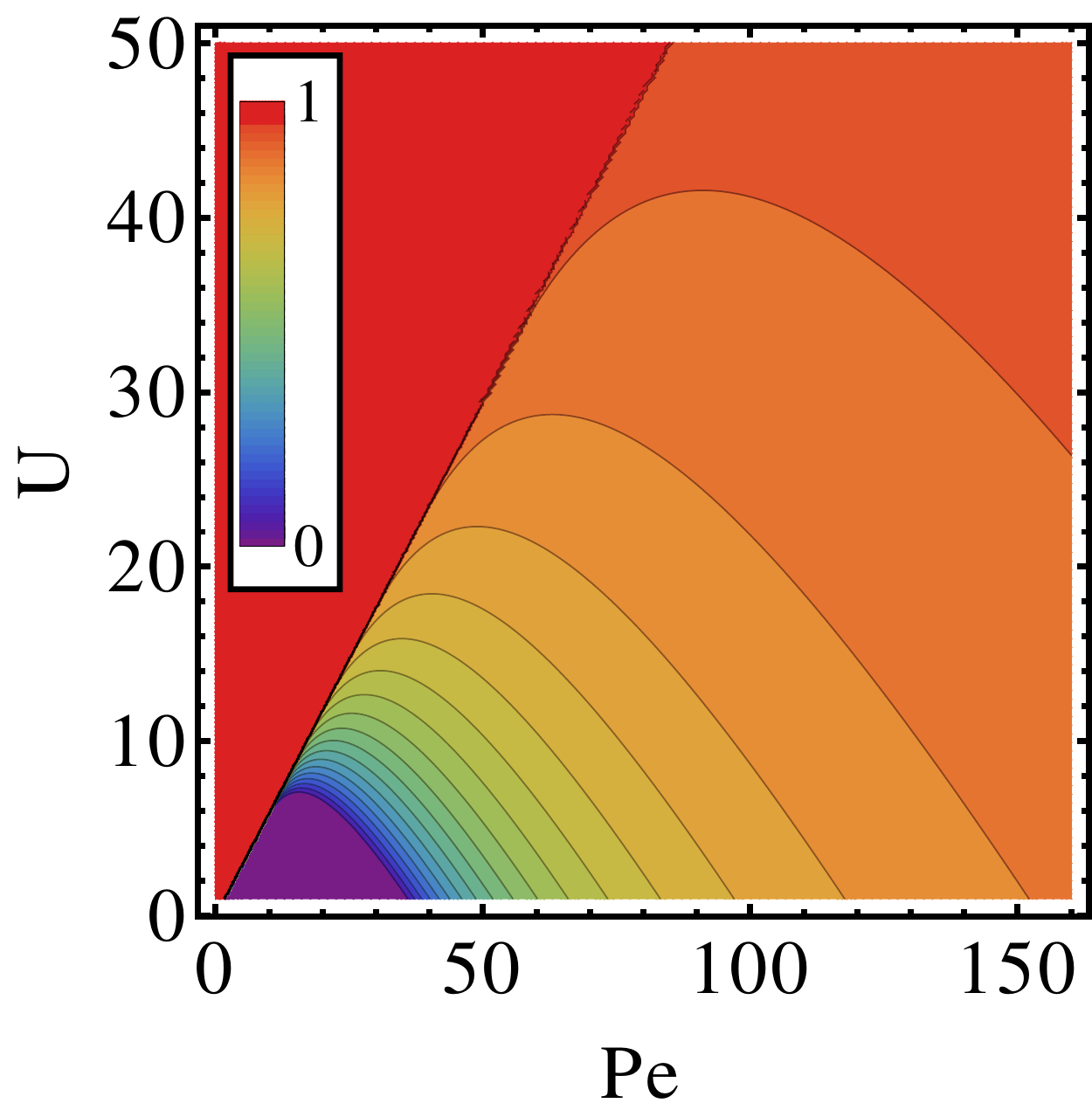}
 \caption{(Color online) {\bf Left:} Fraction of particles in clusters
   $\fc$ measured from simulations as a function of $\Pe$ and $U$.
   {\bf Right:} $\fc$ as predicted by our analytic theory
   (Eq. \ref{eq:cluster_fraction_theory}) which reproduces the major
   features of the phase diagram, including the gel region for $\Pe <
   U$, the self-trapping region for high $\Pe$, and the low-$\fc$
   fluid regime in between.  The values of the adjustable parameters
   are $\kappa = 2$ and $\fmax = 1.7$, with the fit made by eye.}
  \label{fig:phase}
\end{figure}

\begin{figure}[tbp]
  \includegraphics[width=\columnwidth,trim=20 15 0 20,clip]{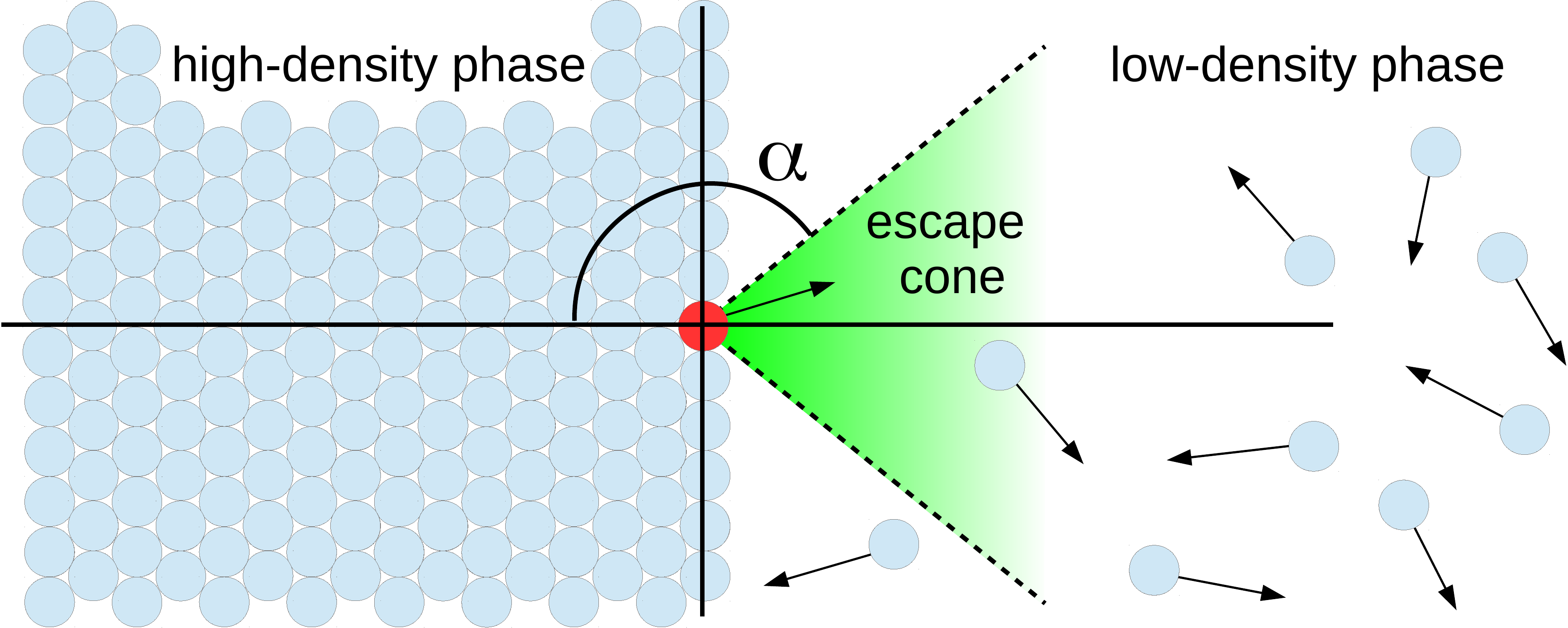}
   \caption{(Color online) Schematic representation of our kinetic
     model.  The high- and low-density phases fill the left and right
     half-spaces.  A particle on the cluster surface (center) escapes
     only if its direction points within the ``escape cone'' defined
     by $\gamma \vp (\nuhat \cdot \nhat) > \Fmax$, while
     particles in the gas land on the surface at a rate proportional
     to the gas density and propulsion speed.}
\label{fig:cartoon}
\end{figure}

To better understand the physical mechanisms underlying the reentrant
phase behavior, we develop a minimal kinetic model to describe the
phase separated state.  By analytically solving the model we obtain a
form for $\fc$ which captures the major features of the phase behavior
observed in our simulations. In the model we consider a single large
close-packed cluster coexisting with a dilute gas which is assumed to
be homogeneous and isotropic (Fig. \ref{fig:cartoon}).  Particles in
the cluster interior are assumed to be held stationary in cages formed
by their neighbors, but their propulsion directions $\theta_i$
continue to evolve diffusively.

To calculate the rate at which gas-phase particles condense onto the
cluster, we observe that the flux of gas particles traveling in a
direction $\nuhat$ through a flat surface is $\frac{1}{2 \pi} \rhog
\vp (\nuhat \cdot \nhat)$, with $\rhog$ the number density of the gas
and $\nhat$ normal to the surface.  Integrating over angles for
particles traveling toward the surface of our cluster yields the
condensation flux $\kin = \frac{\rhog \vp}{\pi}$.

Next we estimate the rate of evaporation.  We note that a particle on
the cluster surface remains bound so long as the component of its
effective propulsion force along the outward normal $\gamma \vp
(\nuhat \cdot \nhat)$ is less than $\Fmax$, the maximum restoring
force exerted on a particle being pulled away from the surface.  This
force may involve multiple bonds and is not simply related to the
interparticle attraction force.  As shown in Fig. \ref{fig:cartoon},
this implies an ``escape cone'' in which the particle's director must
point in order for it to escape.  The critical angle is $\alpha = \pi
- \cos^{-1} \left( \frac{U \fmax}{\Pe} \right)$, with $\fmax$ the
non-dimensionalized maximum restoring force scaled by the depth of the
attractive well, which subsumes all relevant details of the binding
force and is treated as a fitting parameter: $\fmax = \frac{\Fmax}{U}
\frac{\sigma}{\kt}$.

We now consider the steady-state angular probability distribution of
particles on the cluster surface $P(\theta)$.  In the absence of
condensation, this distribution evolves according to the diffusion
equation with absorbing boundaries at the edges of the escape cone:
$\frac{\partial P(\theta, t)}{\partial t} = D_r \frac{\partial^2
  P(\theta, t)}{\partial \theta^2}$ and $P(\pm \alpha, t) = 0$, with
general solution $P(\theta, t) = \sum_{q=1}^\infty A_q \cos \left(
\frac{q \pi \theta}{2 \alpha} \right) e^{-D_r \frac{q^2 \pi^2}{4
    \alpha^2} t}$.  The flux of particles leaving the cluster is then
$\kout = - \frac{1}{\sigma} \left. \frac{\partial}{\partial t}
\int_{-\alpha}^\alpha P(\theta, t) d \theta \right|_{t = 0}$.  To
simplify the analysis we note that higher-order terms decay rapidly in
time, so the steady-state behavior is dominated by the $q=1$ term.  We
therefore discard higher-order terms and solve to find $\kout =
\frac{D_r \pi^2}{4 \sigma \alpha^2}$.

From visual observations it is clear that this minimal model does not
capture all microscopic details of the interfacial region.  In reality
the cluster surface is neither flat nor close packed, but has a
complex form which is constantly reshaped by fluctuations in both
phases (see \ref*{Supplement-fig:interface-high} and
\ref*{Supplement-fig:interface-low} in the SI
\cite{SupplementalInformation}).  We therefore expect quantitative
deviations from the model predictions, which we capture in a general
fitting parameter $\kappa$ which modifies the evaporative flux: $\kout
= \frac{D_r \pi^2 \kappa}{4 \sigma \alpha^2}$.

Equating $\kin$ and $\kout$ yields a steady-state condition which can
be solved for the gas density $\rhog$.  Since the densities of the two
phases are known (the cluster is assumed to be close-packed with
density $\rhoc = \frac{2}{\sigma^2 \sqrt{3}}$) and the number of
particles is fixed, we can calculate the cluster fraction $\fc$:

\begin{equation}  \label{eq:cluster_fraction_theory}
\fc = \frac{16 \phi \alpha^2 \Pe - 3 \pi^4 \kappa}{16 \phi \alpha^2 \Pe - 6 \sqrt{3} \pi^3 \phi \kappa}
\end{equation}

As shown in Fig. \ref{fig:phase}, this model reproduces the essential
features of our system, including active suppression of phase
separation at low $\Pe$, activity-induced phase separation at high
$\Pe$, and a reentrant phase diagram.  The model thus extends the
analysis in Ref. \cite{Redner2013} to describe the coupled effects of
activity and energetic attraction. As noted in that reference, our
model description of self-trapping can be considered a limiting case
of the theory of Tailleur and Cates \cite{Tailleur2008, Cates2013} in
which a self-propulsion velocity that decreases with density leads to
an instability of the homogeneous initial state.

\section{Phase Separation Kinetics}

\begin{figure}[tbp]
  \includegraphics[width=\columnwidth]{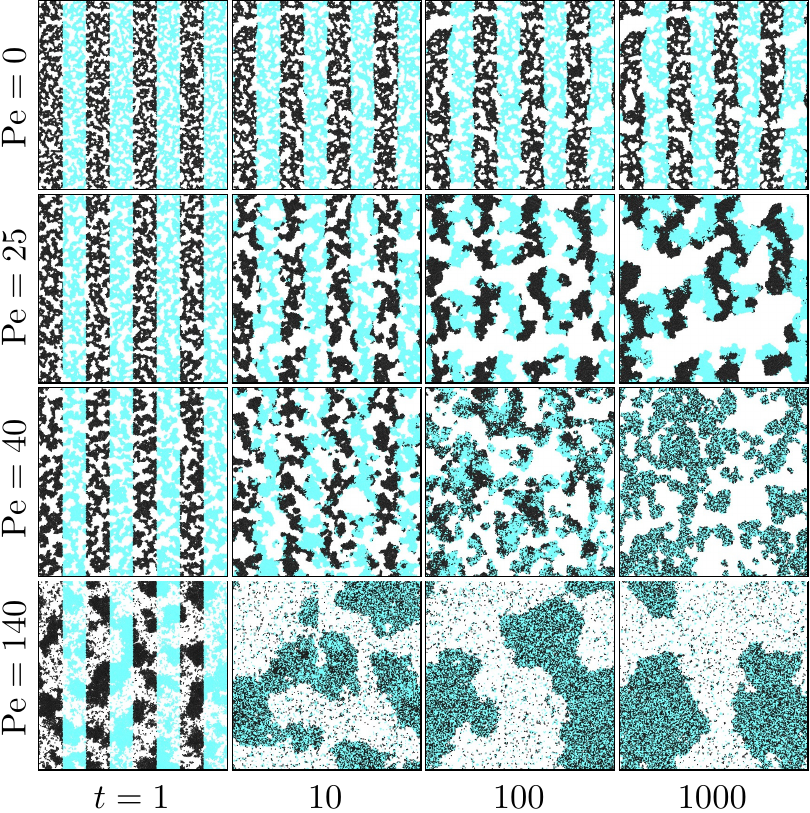}
  \caption{(Color online) Visual guide to phase-separation kinetics at
    fixed $U = 30$.  To make mixing visible, particles are labeled in
    two colors based on their positions at $t=1$.  A passive system
    (top row) forms a space-spanning gel which gradually coarsens.
    The addition of activity (second row) greatly increases the rate
    at which the gel evolves.  In both cases particles remain `local'
    and largely retain the same set of neighbors.  When $\Pe$ exceeds
    $U$ (third row), activity is strong enough to break the gel
    filaments and a fluid of large mobile clusters results.  While the
    instantaneous configurations appear structurally similar to the
    gels above, these systems quickly become well-mixed due to
    splitting and merging of clusters (see also
    Fig. \ref{fig:coarsening}).  In the high-$\Pe$ limit (bottom row),
    self-trapping drives the emergence of a single well-mixed cluster
    surrounded by a dilute gas.}
  \label{fig:kinetics}
\end{figure}

\begin{figure}[tbp]
  \includegraphics[width=.51\columnwidth]{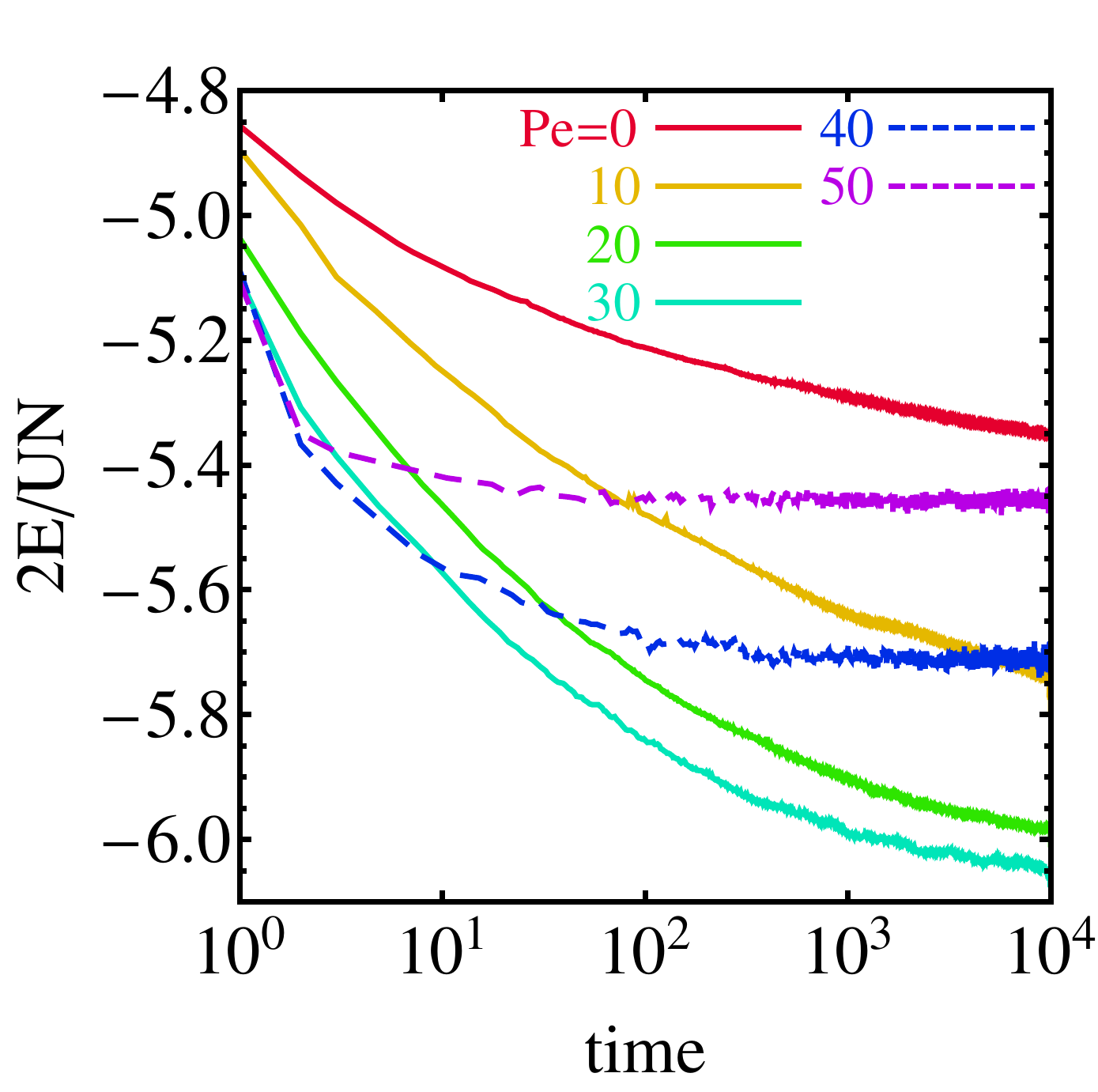}%
  \hfill%
  \includegraphics[width=.485\columnwidth]{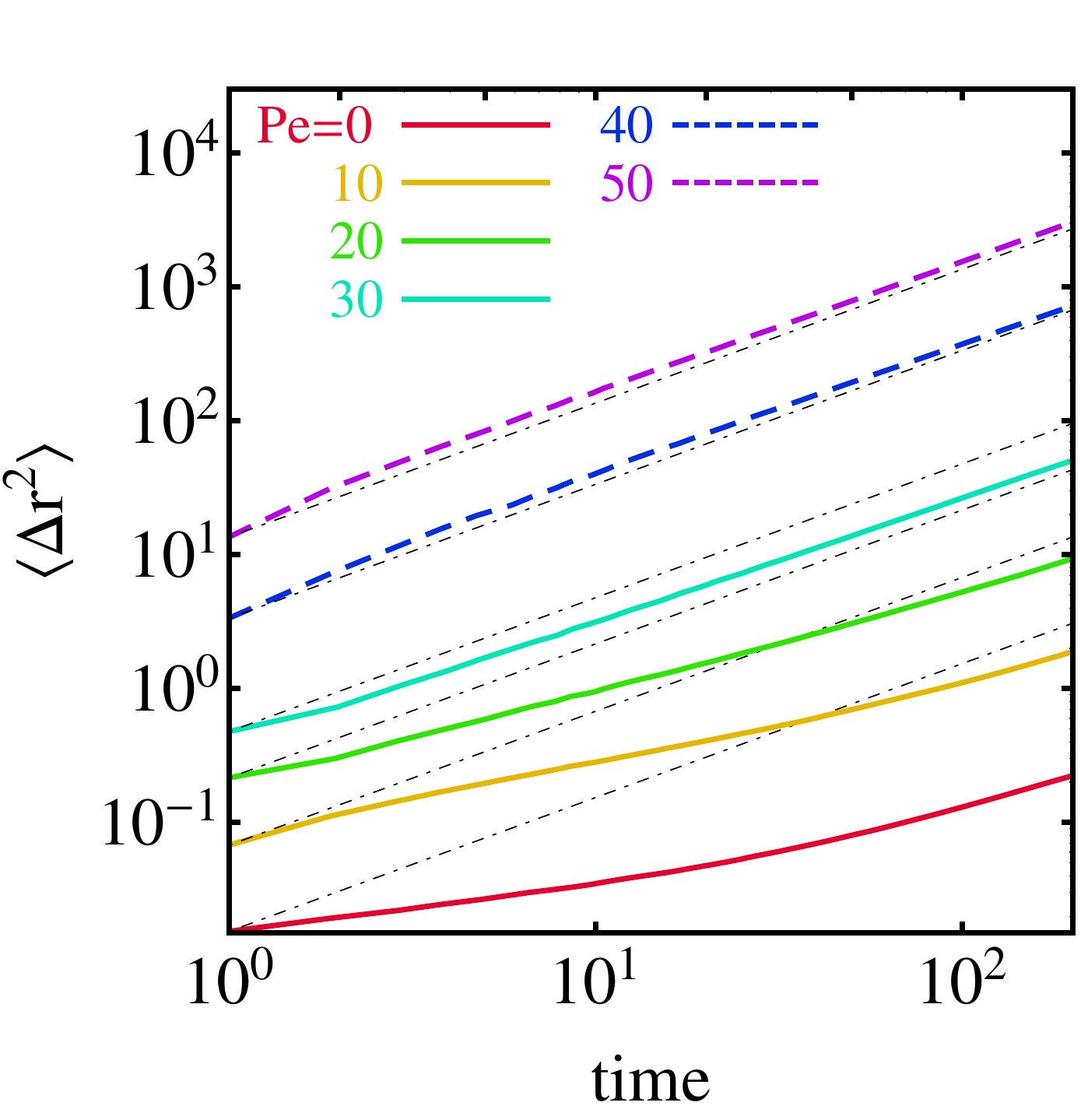}%
  \caption{(Color online) Quantitative measurements of states with
    $U=30$.  {\bf Left: } Mean interaction energy per particle ($E$
    represents the total system potential energy).  Low-$\Pe$ gels are
    strongly arrested and do not approach bulk phase-separation on our
    simulation timescales; however, higher-$\Pe$ systems evolve much
    faster and nearly reach the bulk limit.  When $\Pe > U$ (dashed
    lines), coarsening is arrested as the gel breaks into a fluid of
    mobile clusters with a $\Pe$-dependent characteristic size,
    leading to a plateau in the system potential energy. {\bf Right:}
    Discrimination of gel from fluid states by mean-square
    displacement.  Dotted lines have slope $1$ and highlight the
    distinction at short times between gels (subdiffusive), and active
    fluids (superdiffusive).}
  \label{fig:coarsening}
\end{figure}

The kinetics of phase separation differ significantly between the
low-$\Pe$ gel and high-$\Pe$ self-trapping regions.  In low-$\Pe$
systems, thermal influences dominate and the kinetics are those of a
colloidal particle gel \cite{Trappe2004}.  Since the area fraction in
our simulations is high, the gels we observe appear nonfractal.
Thermal agitation gradually reorganizes the gel into increasingly
dense structures \cite{Poon1998, DArjuzon2003}, leading toward a
single compact cluster in the infinite-time limit.  The presence of
activity greatly accelerates the rate at which the gel evolves, as
shown in Figs. \ref{fig:kinetics} and \ref{fig:coarsening}.  This
effect is also visible in Fig. \ref{fig:overview}, as the apparent
correlation length in the gel states (each observed after a fixed
amount of simulation time) increases with $\Pe$.

As $\Pe$ is increased beyond the threshold value $\Pe \approx U$,
activity begins to overwhelm energetic attraction and the gel is
ripped apart.  This arrests the compaction, resulting in a plateau in
the system's total potential energy (Fig. \ref{fig:coarsening}).  Just
above this transition, the system resembles a fluid of large mobile
clusters which rapidly split, translate, and merge
(Fig. \ref{fig:kinetics}).  As $\Pe$ is further increased, the
characteristic mobile cluster size decreases until the appearance of
an ordinary active fluid of free particles is recovered.  The fluid
phase is clearly identified by superdiffusive mean-square displacement
measurements (Fig. \ref{fig:coarsening}), distinct from the
subdiffusive behavior found in gels.

Additional increase of $\Pe$ will eventually cross a second threshold
into a phase-separated regime whose behavior is dominated by
self-trapping.  As reported previously \cite{Redner2013}, these
systems undergo nucleation, growth, and coarsening stages in a manner
familiar from the kinetics of quenched fluid systems, albeit with the
unfamiliar control parameter $\Pe$ instead of temperature.

These three regimes are characterized by dramatically different
particle dynamics. To illustrate these behaviors and their effect on
particle reorganization timescales, in Fig.~\ref{fig:kinetics} we
present snapshots from simulations in which initial particle positions
are identified by color. We see that when attraction is dominant
($U>\Pe$), each particle's set of neighbors remains nearly static over
the timescales simulated, indicative of long relaxation timescales due
to kinetic arrest. In contrast, when activity dominates ($\Pe>U$)
particles rapidly exchange neighbors and the system becomes
well-mixed. Importantly, note that particle dynamics cannot be
directly inferred from the instantaneous spatial structures in the
system. For example, while systems with low activity ($\Pe<U$,
Fig.~\ref{fig:kinetics} second row) and moderate activity ($\Pe
\gtrsim U$, Fig.~\ref{fig:kinetics} third row) appear structurally
similar, the rate of particle mixing differs by orders of magnitude.

\section{Conclusions}

Activity can both suppress and induce phase separation, and we have
shown that these opposing effects can coexist in the same simple
system.  The resulting counterpoint produces a reentrant phase diagram
in which two distinct types of phase separation exist, separated by a
homogeneous fluid regime.  This surprising result makes it possible to
use two experimentally accessible control parameters ($\Pe$ and $U$)
in concert to tune the phase behavior of active suspensions.  This
control is especially valuable because attractive interparticle
interactions are common in experimental active systems, being either
intrinsic \cite{Theurkauff2012a, Palacci2013} or easily imposed, such
as by the addition of depletants \cite{Schwarz-Linek2012}.  An
understanding of the complex phase behavior accessible to these
systems is a critical stepping stone toward designing smart active
materials whose phases and structural properties can dynamically
respond to conditions around them.

\section{Acknowledgments}

This work was supported by NSF-MRSEC-0820492 (GSR, AB, MFH), as well
as NSF-DMR-1149266 (AB). Computational support was provided by the
National Science Foundation through XSEDE computing resources
(Trestles) and the Brandeis HPC.

\bibliographystyle{apsrev4-1}
\bibliography{paper}

\end{document}